\documentclass[iop]{emulateapj}
\begin{document}

\title{On the Relative Sizes of Planets within $Kepler$
Multiple Candidate Systems}

\author{David R. Ciardi\altaffilmark{1},
Daniel C. Fabrycky\altaffilmark{2},
Eric B. Ford\altaffilmark{3},
T. N. Gautier III\altaffilmark{4},
Steve B. Howell\altaffilmark{5},
Jack J. Lissauer\altaffilmark{5},
Darin Ragozzine\altaffilmark{3},
Jason F. Rowe\altaffilmark{5}}
                             
\altaffiltext{1}{NASA Exoplanet Science Institute/Caltech Pasadena,  CA
91125 USA}
\altaffiltext{2}{UCO/Lick Observatory, U. of California, Santa Cruz, CA,
USA}
\altaffiltext{3}{Dept. of Astronomy, University of Florida, Gainesville,
FL, USA}
\altaffiltext{4}{Jet Propulsion Laboratory, Pasadena, CA, USA}
\altaffiltext{5}{NASA Ames Research Center, Mountain View, CA USA}

\email{ciardi@ipac.caltech.edu}



\slugcomment{Accepted for publication in The Astrophysical Journal}

\begin{abstract} 

We present a study of the relative sizes of planets within the multiple
candidate systems discovered with the $Kepler$ mission.  We have compared
the size of each planet to the size of every other planet within a given
planetary system after correcting the sample for detection and  geometric
biases. We find that for planet-pairs for which one or both objects is 
approximately Neptune-sized or larger, the larger planet is most often the
planet with the longer period.  No such size--location  correlation is seen
for pairs of planets when both planets are smaller than Neptune. 
Specifically, if at least one planet in a planet-pair has a radius of
$\gtrsim 3R_\oplus$,  $68\pm 6\%$ of the planet pairs have the inner planet
smaller than the outer planet, while no preferred sequential ordering of
the planets is observed if both planets in a pair are smaller than
$\lesssim3 R_\oplus$. 

\end{abstract}

\keywords{planetary systems}

\section{Introduction}\label{intro-sec}

Approximately 20\% of the planetary candidate systems discovered thus far
by $Kepler$ \citep{borucki10} have been identified as having multiple
transiting candidate planets \citep{batalha12}.  As $Kepler$ continues its
mission, the number of multiple planet systems is likely to grow - not only
because the total number of systems known to host planets will increase,
but also because previously identified ``single''-candidate systems may be
found to have additional planets not previously detected. For example, 96
(11\%) of the ``single'' candidate systems from the \citet{borucki11}
$Kepler$ Object of Interest (KOI) list are now listed on the
\citet{batalha12} KOI list as having multiple planetary candidates. Thus,
understanding planets in multiple systems is not only important for placing
into context our own Solar System (the best studied planetary system), but
multiple planetary systems may turn out to be the rule rather than the
exception.  

Typical detailed confirmation of individual planetary systems takes a
concerted ground-based observational and modeling effort to rule out false
positives caused by blends \citep[e.g., ][]{batalha11, torres11}.  However,
the multiple candidate systems provide additional information that can be
used to confirm planets directly via transit timing variations
\citep{steffen12a, fabrycky12a} or via statistical arguments that multiple
transiting systems discovered by $Kepler$ are almost all true planetary
systems \citep{latham11, lissauer11b, lissauer12}.  Indeed,  based upon
orbital stability arguments, $\approx 96\%$ of the pairs within multiple
candidate systems are most likely real planets around the same star
\citep{fabrycky12b}.

The higher statistical likelihood that candidates in multiple systems are
true planetary systems has enabled studies of the properties of  planetary
systems with a lower level of false positive contamination than would be
expected if all transiting systems were studied.  The overall false
positive rate within the $Kepler$ candidate sample has been estimated to
be  $10 - 35$\% \citep[e.g.,][]{mj11,santerne12}, but in comparison, the 
overall false positive rates among the multiple transiting systems is
expected to be $\lesssim 1$\% \citep{lissauer12}.

The multiple candidate systems also provide another level of certainty to
the studies of planetary systems.  The quality of knowledge of planetary
characteristics  for individual planets often is dependent upon the quality
of knowledge of the host stellar characteristics.  For example, planetary
radii uncertainties for transiting planets are dominated by the
uncertainties in the stellar radii $({\rm transit\ depth} \propto
(R_p/R_\star)^2)$, and changes in our understanding of the stellar radius
can greatly alter our understanding of the radii of individual planets
\citep[e.g.,][]{muirhead12}.  By studying the {\it relative} properties of
planets within multiple systems, systematic uncertainties associated with
the stellar properties are minimized,   

Understanding the relative sizes and, hence, the relative bulk compositions
and structures of the planets within a system can yield clues on the
formation, migration, and evolution of planets within an individual system
and on planetary systems as a whole \citep[e.g.,][]{raymond12}.    Within
our Solar System, the distribution of the unique pair-wise radii ratios for
each planet compared to the planets in orbits exterior to its orbit  (e.g.,
Mercury-Venus, Mercury-Earth, $\ldots$ Mercury-Neptune, Venus-Earth,
Venus-Mars, $\ldots$ etc.) is dominated by ratios less than unity (i.e.,
the inner planets are smaller than the  outer planets).  For the 8 planets
in the Solar System, 20 of the 28 ($70\%$) unique radii ratios are $<1$,
but only 3 out of 7 ($42\%$) of neighboring pairs display this sequential
size  hierarchy.  If only the terrestrial planets are considered (Mercury,
Venus, Earth, Mars), the fraction is $2/3$ with only Mars being smaller
than its inner companions.  

How the planets sizes are ordered and distributed is likely a direct result
of the how the planets formed and evolved.  For example, Mars, being the
only terrestrial planet in our Solar System that does not follow the
sequential planet size distribution, may be a direct result of the 
formation and migration of Jupiter, inwards and then back outwards leaving
a truncated and depleted inner disk out of which Mars was formed
\citep{walsh11}.

\citet{lissauer11b} noted, for adjacent planets within the $Kepler$
multiple-candidate systems, that the distribution of radii ratios for
adjacent planets is symmetrically centered around unity, but hints of a
planetary size hierarchy can be seen in multi-planet systems discovered by
Kepler.  $Kepler$-11, which has 6 transiting planets  \citep{lissauer11a}
with the smallest planets residing preferentially inside the orbits of
larger planets.  $Kepler$-11 displays an anti-correlation between the mean
density of the planets and the semi-major axis distance from the host star;
i.e., the larger and lower density planets are located further out than the
smaller and denser planets, possibly indicative of the formation and/or
evolution of the planetary system \citep{msg12}. $Kepler$-47, the only
known circumbinary multiple planet system, also displays a size hierarchy
with the inner planet ($\sim 3R_\oplus$) being smaller than the outer
planet \citep[$\sim4.6R_\oplus$;][]{orosz12}. Yet, in $Kepler$-20 (a
5-planet system), the relative sizes of the planets do not appear to
correlate with the orbital periods of the planets \citep{gautier12,
fressin12}.  But these are only three systems.  Is there an overall
correlation of planetary size with orbital period?

Here we explore the {\it relative} sizes of the planetary radii for all
planet pairs within the multiple candidate systems discovered thus far by
$Kepler$ \citep{batalha12}.  In this work, we refer to the Kepler
candidates as ``planets'' though the majority have not been formally
validated or confirmed as planets; as discussed above, candidates in
multiple candidate systems are statistically more likely to be true
planetary systems \citep{latham11, lissauer11b, lissauer12}. We seek to
characterize the planetary size hierarchy, as a function of the number of
planets detected in the systems and the properties of the stellar hosts,
and explore if the size hierarchy seen within the inner Solar System also
occurs in the $Kepler$ multiple-candidate sample.

\section{The Sample}\label{sample-sec}

The sample used here is based upon the 2012 KOI candidate list published by
\citet{batalha12}; a full description of the KOI list, the  vetting that
list underwent, and the characteristics of the sample set as a whole are
described in the catalog paper.  There are 1425 systems with a single
candidate, 245 systems with two planet candidates, 84 systems with three
planet candidates, 27 systems with four planet candidates, 8 systems with
five planet candidates and one system with 6 planet candidates. Here we
wish to investigate the overall distribution of planet sizes as a function
of orbit; that is, do outer planets, in general, tend to be larger than
inner planets?
\begin{figure}[htp]
	\centering
    \includegraphics[angle=0,scale=0.5,keepaspectratio=true]{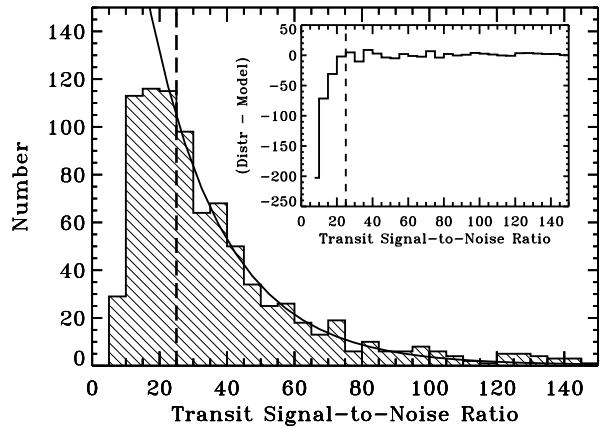}

    \figcaption{Distribution of the transit signal-to-noise ratio for all
    the detected ``multiple'' candidates from \citet{batalha12}.  The 
    solid curve is an exponential fit to the SNR distribution, and the 
    vertical dashed line at $SNR=25$ marks approximately where the
    exponential no  longer adequately describes the distribution (see inset
    figure).\label{snr-fig}}

\end{figure}

A smaller planet (e.g., shallower transit) will be more easily
detected if the period is short, simply from the fact that the number of
observed transiting events increases with shorter orbital period. In a
complementary manner, the larger planets are more easily detected at all
periods, up to some period threshold where only $2-3$ transits are
detected.  \citet{batalha12} tabulate for each candidate the transit
period, the  transit duration, the impact parameter, and the
signal-to-noise of the transit fits.  The signal-to-noise ratios, coupled
with the transit periods and the impact parameters,  enables us to debias
the sample for these observational detection efficiencies \citep[see
also][]{lissauer11b}.

The multiple candidate systems were identified primarily by searching the
single candidate systems specifically for additional transiting planets. 
Thus, there could potentially be a detection bias of which we are unaware
that is not found in the single candidate systems. However, no substantial
difference between single planet systems and the multiple planet systems
were found, except for the lack of hot Jupiters in the multiple planet
systems \citep{latham11,steffen12b}.  The overall size distribution of the
planets and the signal-to-noise detection thresholds appear to be similar
between the  single candidate and multiple candidate systems.
\begin{figure}[h]
	\centering
    \includegraphics[angle=0,scale=0.5,keepaspectratio=true]{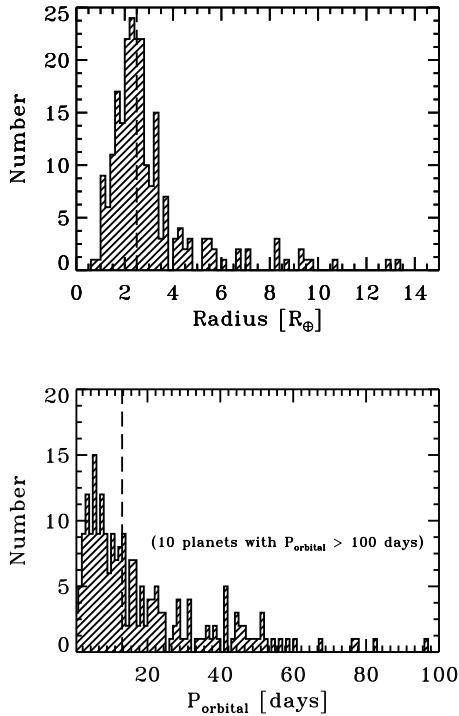}

    \figcaption{Distributions of the radii and orbital periods of the
    planets that are used in this study.  The vertical dashed lines
    mark the median values of the distributions.\label{planets-fig}}

\end{figure}

To understand better the detection thresholds for the multiple candidate
systems, we have used the distribution of the total transit signal-to-noise
to estimate where the  the distribution of planets in multiple systems
appears to be complete  (Figure~\ref{snr-fig}).   Assuming the
signal-to-noise distribution can be characterized with an exponential
function where the distribution is complete, we have used the point in the
signal-to-noise  distribution function where the exponential no longer
adequately describes the distribution as the fiducial for completeness. 
Fitting the exponential, we find a turnover in detected samples at a $SNR
\approx 25$.  We use this SNR threshold to debias the radii-ratio
distribution of planets in multiple-candidate systems.

The total signal-to-noise of all the detected transits for each planetary
candidate is predicted for all the observed orbital periods within a
system. A planet-pair is retained in the analysis only if the predicted
total transit signal-to-noise ratios for both planets at the orbital period
of the other planet exceeds the signal-to-noise threshold of $SNR > 25$. 
The predicted signal-to-noise ratio is determined by scaling the measured
total signal-to-noise ratio of the planet by the ratio of the orbital
periods.   Assuming all else is equal, the total signal-to-noise of all the
detected transits for a given planet scales with the orbital period as
$P^{-1/3}$.  

For a given planet, the total signal-to-noise of all detected transits is
proportional to the total number of points ($n$) detected in all of the
transits \citep[e.g.,][]{vbkc09}
\begin{equation} 
	(S/N)_{transits} \propto (n)^{1/2} \propto \left(N \cdot n_t\right)^{1/2}
\end{equation}
where $N$ is the total number of individual transits detected, and $n_t$ is
the number of points detected within a single transit.  The number of transits
detected is inversely proportional to the orbital period $(N \propto
1/P)$, and the  number of points detected per transit is proportional to
the transit duration $(n_t \propto t_{dur})$; thus, the total signal to noise 
of the detected transits can be  parameterized as
\begin{equation}
    	(S/N)_{transits} \propto \left(t_{dur}/P\right)^{1/2}
\end{equation}
The transit duration ($t_{dur}$) is proportional to the cube root of the
orbital period ($t_{dur} \propto v_{orb}^{-1} \propto P^{1/3}$); thus, the
total signal-to-noise ratio of the detected transits scales with the
orbital period as 

\begin{equation}
(S/N)_{transits} \propto (P^{1/3}/P)^{1/2} \propto P^{-1/3}.  
\end{equation}

An additional restriction on the sample was made such that no planet
candidate was included that has an impact parameter of $b \ge 0.8$. At such
high impact parameters, the transit parameters, particularly the transit
depth (i.e., the planet radius), are less certain.   Using the $SNR > 25$
detection threshold and the impact parameter restrictions, there are 96
multiple-candidate systems with 159 pairs of planets (228 individual
planets) in the analysis.    

All of the planet-pairs used in the analysis of this paper are summarized
in Table~\ref{sum-tab}; planets are labeled with roman numerals
(I,II,III,IV,V,VI) in the order of increasing orbital period. These letters
do not necessarily correspond to KOI fraction numbers (e.g., .01, .02, ...)
nor do they correspond to the confirmed planet letters (e.g., $Kepler$-11b,
$Kepler$-11c, ...). After the signal-to-noise and impact parameter cuts,
the largest planet in the sample is $13 R_{\oplus}$, and the median planet
radius is $\approx 2.5 R_{\oplus}$; the smallest planet retained in the
sample has a radius of  $0.75 R_{\oplus}$.  The orbital periods of the
planets in this sample span 0.45 days to 331 days, with an median period of
$\sim 13.1$ days.  The distributions of the radii and orbit periods for the
228 planets retained in the sample are shown in Figure~\ref{planets-fig}.

Previous work indicates that the detected planets in the $Kepler$
multiple-candidate systems have mutual inclinations of $1^\circ - 3^\circ$
\citep{fabrycky12b,fm12}.  Due to the usual limitation of the transit
technique in only detecting planets that are very nearly edge-on, the
Kepler sample used here is naturally biased to systems of nearly coplanar
planets. The prevalence of Kepler multiple-candidate systems shows that
there is a large population of such systems with small planets and orbital
periods of tens of days (Figure~\ref{planets-fig}). It may be that other
system architectures with higher mutual inclinations or different period
ranges do not show the same trend in planet sizes that we describe herein. 

\begin{figure}[htp]
	\centering
    \includegraphics[angle=0,scale=0.5,keepaspectratio=true]{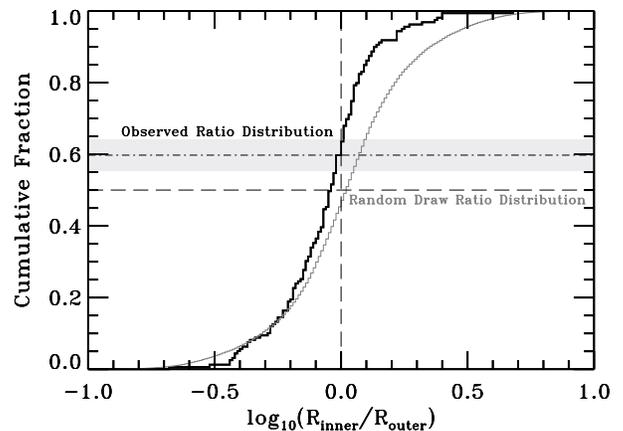}

    \figcaption{Observed cumulative distribution of the planet-radii
    ratios  for all planet pairs (black histogram), and the predicted
    cumulative distribution for planet radii drawn randomly the measure
    planet sample (grey histogram).  The horizontal dot-dash line marks the
    fraction of planet pairs with $R_{inner}/R_{outer} < 1$;  the grey
    region marks the $1\sigma$ confidence interval for this fraction.  The
    vertical dashed line marks the boundary where $R_{inner}/R_{outer} =
    1$, and the horizontal dashed line marks the 50\%
    fraction.\label{ratiocfrac-fig}}

\end{figure}
\section{Discussion}\label{discussion-sec}

We have calculated the ratios of the inner planet radius to the outer
planet radius for each unique pair of planets within a system
($R_{inner}/R_{outer}$), and the cumulative fraction distribution is
displayed in Figure~\ref{ratiocfrac-fig}.  If there were no preference for
the ordering of planet sizes, the chance that a given planet-radii ratio is
larger than unity would be equal to the chance that the ratio is below
unity, and the cumulative fraction distribution would pass through 50\% at
$\log(R_{inner}/R_{outer}) = 0\ (R_{inner}/R_{outer}=1)$. 

In contrast, $59.7^{+4.1}_{-4.2}$\% of the planet pairs are ordered such
that the outer planet is larger than the inner planet
($R_{inner}/R_{outer}<1$). The resulting fraction deviates from the
null-hypothesis expectation value of 50\% by $\approx 2.5\sigma$.  The
$1\sigma$ upper and lower confidence intervals are based upon the
Clopper-Pearson binomial distribution confidence interval \citep{cp34}. 
The Clopper-Pearson interval is a two-sided confidence interval and is
based directly on the binomial distribution rather than an approximation
to the binomial distribution.  We do caution that in applying the
Clopper-Pearson confidence interval there is an implicit assumption that
all of elements of the sample are uncorrelated.  Given that not all
planet-pairs are from independent stellar systems, there  may be a
correlation between individual planet-pairs within a system (i.e., planet I
is smaller than planet III because it is smaller than planet II), and the
Clopper-Pearson confidence intervals may underestimate the true
uncertainties in the fractions.

In addition to the confidence intervals, the significance of the observed
fraction in comparison to the null hypothesis can be evaluated through the
$\chi^2$ statistic.  The probability of the observed fraction ($96/159 =
59.7\%$) when compared to the expected fraction ($79.5/79.5 = 50\%$) yields
$\chi^2$ = 6.0  with a probability of only 1.4\% that the observed
fraction is observed only by chance (see Table~\ref{frac-tab}).

We also have performed three additional observational tests to assess if
the observed fraction may be the result of an unrecognized bias in the
sample.  The first test repeated the above analysis, but for each unique
planet pair, the measured radii were replaced with radii randomly drawn
from the overall radius distribution as measured by Kepler; the same set of
periods within a given system were retained.  The same impact parameter and
signal-to-noise restrictions described above (\S\ref{sample-sec}) were
applied (e.g., if a planet radius was drawn that was too small to be
detected at the orbital period ($SNR<25$), a new radius was randomly drawn
until the $SNR$ threshold was met).  The random draw was performed 10,000
times for each unique pair of planets, and the cumulative distribution of
the planet-radii ratios for all of the random draws is displayed in
Figure~\ref{ratiocfrac-fig}.  As expected, the random draw distribution
displays no size ordering preference; i.e., the fraction of planets with
$R_{inner}/R_{outer}<1$ is $\approx 50\%$.  Based upon a 
Kolmogorov-Smirnov test, the observed distribution and the random draw
distribution are not drawn from different parent distributions with only a
probability of $3\times 10^{-8}$ -- indicating that the observed planet
radii ratio distribution has a preferential ordering such that smaller
planets are in orbits interior to larger planets.
\begin{figure}[htp]
	\centering%

    \includegraphics[angle=0,scale=0.5,keepaspectratio=true]{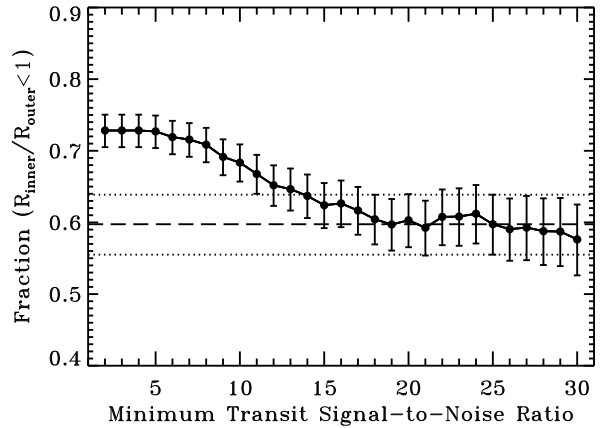}

    \figcaption{The observed fraction of planet pairs with 
    $R_{inner}/R_{outer} < 1$ is plotted as a function of signal-to-noise
    cut-off, showing the fraction asymptotically approaches the a value of
    $\sim 0.6$ for $SNR \gtrsim 20$.  The dashed line marks the observed
    fraction (and the $1\sigma$ confidence interval; dotted lines) for a
    $SNR_{threshold} = 25$ (from
    Fig.~\ref{ratiocfrac-fig}).\label{snrratio-fig}}

\end{figure}

We also tested if the observed fraction of planet-radii ratios with
$R_{inner}/R_{outer}<1$ is dependent upon the signal-to-noise ratio
threshold chosen.  In Figure~\ref{snrratio-fig}, the fraction of  planet
pairs where the inner planet is smaller than the outer planet is plotted as
a function of the required signal-to-noise threshold.  If no
signal-to-noise threshold is required, the fraction is $>70\%$, and as the
signal-to-noise threshold is increased, the fraction systematically
decreases and levels out near $\sim 60\%$.  The higher fractions at a lower
signal-to-noise threshold result from incompleteness of the sample
(i.e., it is easier to detect larger planets at all orbital periods).  As
the signal-to-noise threshold is increased the fraction of planet-radii
ratios below unity decreases, but does not systematically approach 50\%, as
would be expected if there was no preferential size ordering of the
planets.  Rather, the fraction asymptotically approaches 60\% for
$SNR_{threshold} > 20$, indicating the $SNR_{threshold}=25$ threshold
ensures that the analysis presented in Fig.~\ref{ratiocfrac-fig} is based
upon a sample not significantly biased by completeness.
\begin{figure}[htp]
	\centering%

    \includegraphics[angle=0,scale=0.5,keepaspectratio=true]{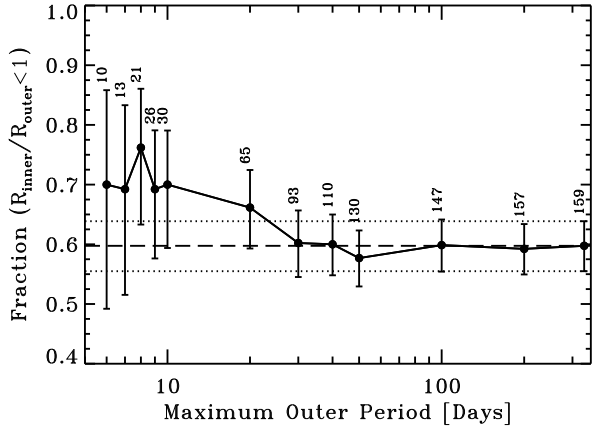}

    \figcaption{The observed fraction of planet pairs with 
    $R_{inner}/R_{outer} < 1$ is plotted as a function of maximum outer
    orbital period cut-off, showing the fraction asymptotically approaches
    the value of $\sim 0.6$ for $P \gtrsim 20$.  The dashed line marks the
    observed fraction for the whole sample; the dotted lines mark the
    $1\sigma$ confidence interval (from Fig.~\ref{ratiocfrac-fig}).  The
    numbers above each data point indicate the number of planet-pairs that
    appear in that period bin ($P_{outer} <
    P_{max}$).\label{ratioperiod-fig}}

\end{figure}

The final test performed was to determine if the observed fraction is
dependent upon the maximum period of the outer planet (see
Figure~\ref{ratioperiod-fig}).  Transit surveys are typically more complete
at shorter orbit periods, and if there truly was no preference for planet
size ordering, the fraction would decline and approach 50\% as the maximum
orbit period was decreased.  Within the confidence intervals, the observed
fractions are independent of the maximum outer orbital period and are
consistent with the observed $\approx 60\%$ fraction for the entire sample.
The fractions do not approach 50\%, at shorter orbital period as would be
expected for a random  distribution. In fact, a slight (but not
statistically significant) hint of a higher fraction is observed if the
maximum period is $P < 20$ days.

Overall, if the ordering of planets within a given system was random such
that there was no sequential ordering of the planets by size, the
probabilities of a planet-radii ratio being above or below unity would be
equal. But that is not what is observed; the three above tests (i.e., the
random radius draw, the $SNR_{threshold}$, and the maximum outer orbital
period), in combination with the $\chi^2$ probability statistic and the
confidence intervals, indicate that for $\approx 60\%$ ($2.5\sigma$) of the
unique planet pairs within the Kepler multiple planet systems, the inner
planet is smaller than the outer planet, with only a 1.4\% chance of this
being observed by chance.  In the following subsections, we explore if and
how the number of planets in the system, the orbital separation of the
planets, the temperature of the stars, and the size of the planets
themselves may affect the observed size hierarchy of the  planets in
multiple-planet systems.

\begin{figure}[htp]
\centering%
    \includegraphics[angle=0,scale=0.5,keepaspectratio=true]{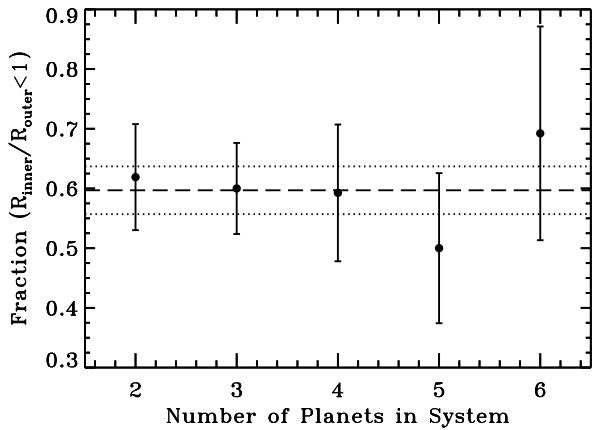}

    \figcaption{Observed fraction of planet-pairs with the inner planet
    being smaller the outer planet plotted as a function of the number of
    KOIs within a system.The dashed line marks the observed fraction for
    the whole sample; the dotted lines mark the $1\sigma$ confidence
    interval (from Fig.~\ref{ratiocfrac-fig}). \label{fracnum-fig}}

\end{figure}
\subsection{The Number of Planets and Orbital Separation}\label{numandorbit-sec}

If formation and evolutionary mechanisms within a planetary system affect
the size distribution and ordering of planets within a system, one might
expect to observe differences in the size ordering as a function of the
number of planets within a system. When the planets are divided into groups
based upon the number of detected planetary candidates in the system (e.g.,
2-, 3-, 4-, 5-, and 6-candidates),  the fractions do not change appreciably
from the fractions calculated for the entire sample set.  This can be seen
in the both $\chi^2$ statistic (Table~\ref{frac-tab}) where the $\chi^2$
and associated probabilities are all comparable to each other and in
Figure~\ref{fracnum-fig}, where the observed fractions (with confidence
intervals) are plotted and display no dependence on the number of planets
in the system.  While the number of systems with more than 3 planets have
relatively large uncertainties and poor statistics because of the small
numbers of the 4-, 5-, and 6-candidate systems, there is no correlation of
the fraction of size-ordered planets with the number of planets within a
planetary system.  

\begin{figure}[h]
	\centering%
    \includegraphics[angle=0,scale=0.5,keepaspectratio=true]{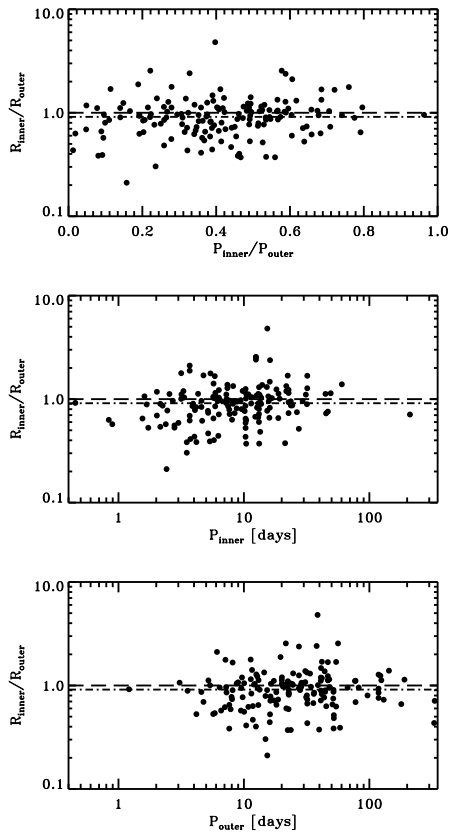}

    \figcaption{{\it Top:} Distribution of the planet-radii ratios as a
    function of the orbital period ratio. {\it Middle:} Distribution of the
    planet-radii ratio as a function of the inner planet orbital  period. 
    {\it Bottom:} Distribution of the planet-radii ratio as  a function of
    the outer planet orbital period. In each panel, the horizontal dashed
    line marks unity, and the dot-dashed line delineates the median
    planet-radii ratio of 0.91 for the entire sample. 
    \label{periodratio-fig}}

\end{figure}

It might be expected that planets nearer to the central star would undergo
a higher level of photo-evaporation than planets at longer orbital
periods and, thus, the relative sizes of the planets would be more extreme
if the planets are more widely separated (larger orbital period ratio).  To
test this, we have plotted the planet-radii ratios vs. the orbital period
ratios and the inner and outer planet orbital periods
(Fig.~\ref{periodratio-fig}).  Using the Spearman non-parametric rank
correlation function, we find that the distributions are likely
uncorrelated. The correlation coefficients for the three distributions
shown in Fig.~\ref{periodratio-fig} are 0.14, 0.22, and 0.10 with
probabilities to be exceeded in the null hypothesis of 0.06, 0.004 and
0.08, respectively.

Additionally, to search for a non-zero slope that might indicate the
planet-radii ratios are related to or dependent upon the period or period
spacings, we have fitted a linear model\footnote{Fitting was done with an
outlier resistant linear regression routine based upon Numerical Recipes
\citep{nr07}.} to each of the distributions. Each of the distributions are
statistically consistent with a flat distribution as a function of period
ratio and orbital period; the slopes of the fitted linear models were found
to be $0.29\pm0.15$ for the radii ratios vs. period ratios
(Fig.~\ref{periodratio-fig} top), $0.0009\pm0.001$ for the radii ratios vs.
the inner orbital period (Fig.~\ref{periodratio-fig} middle), and 
$-0.0002\pm0.0005$ for the radii ratios vs. the outer orbital period
(Fig.~\ref{periodratio-fig} bottom).  In general, we find no correlation
between the orbital period separation (or orbital periods themselves) and
the size-ordering of the planets.

\begin{figure}[h]
	\centering%
    \includegraphics[angle=0,scale=0.5,keepaspectratio=true]{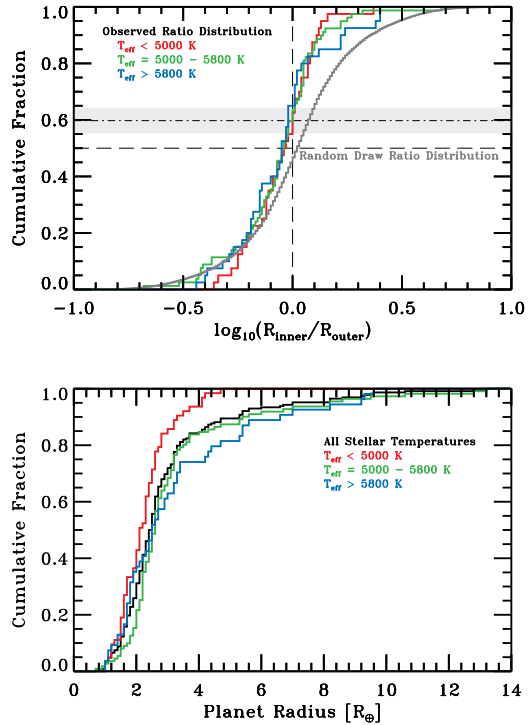}

    \figcaption{{\it Top:} Observed cumulative distributions of the
    planet-radii ratios as is displayed in Fig.~\ref{ratiocfrac-fig},
    except the distributions have been separated out by stellar effective
    temperature and identified by the plot colors (as labeled in the
    plot).   The plot markings as the same as those in
    Fig.~\ref{ratiocfrac-fig}. {\it Bottom:} Cumulative distributions of
    the individual radii for the planets that are used to determine the
    planet-radii ratios throughout the paper, separated out by stellar
    effective temperature (as labeled in the plot). \label{radiiteff-fig}}

\end{figure}

\subsection{Stellar Temperature}\label{teff-sec}

We have also explored whether there is a correlation between the
distribution of planet-radii ratios and the effective temperature  of the
host stars.  We have produced planet-radii ratio distributions, but
separated out by the stellar temperature.  We chose temperature ranges that
roughly correspond to the spectral classification of M and K-stars ($<5000$
K),  G-stars ($5000-5800$ K), and F-stars \citep[$>5800$ K;][]{crd11}.  The
cumulative distributions of the planet-radii ratios for each of the stellar
temperature groups are displayed in Figure~\ref{radiiteff-fig}.

Overall, all three distributions are shifted to lower ratios in comparison
to the random draw distribution from Fig~\ref{ratiocfrac-fig}.
Kolmogorov-Smirnov tests between the three distributions indicate that the
three distributions do not result from different parent distributions with
probabilities of $>75\%$ - indicating that, in general, the preferential
planet size ordering occurs at approximately the same level (i.e., $\approx
60\%$  the unique planet pairs have $R_{inner}/R_{outer} < $) across the
stellar temperature range.  A summary of the planet-radii ratios, the
fraction of ratios that exhibit smaller inner planets, and the significance
of those fractions based upon the confidence intervals and $\chi^2$
statistic is given in Table~\ref{fracstellar-tab} for each of the stellar
temperature groups.  

There does appear to be an slight trend such that as the stars become
warmer they contain a larger fraction of planet-pairs sequentially ordered
by radius.   For the cooler stars ($<5000$ K), this fraction is $55\pm9\%$,
while for the warmer stars ($>5800$ K), this fraction is $65\pm9$\%.  The
resulting $\chi^2$ statistics for these three samples indicates that the
observed fractions differ from the null hypothesis fraction of 50\% more
significantly for the warmer stars than for the cooler stars
(Table~\ref{fracstellar-tab}). This trend could be a result of higher
planetary evaporation rates associated with the higher luminosity stars or
perhaps could result from the warmer (i.e., more massive) stars tending to
contain larger (i.e., more massive) planets (see Figure~\ref{radiiteff-fig}
and Table~\ref{fracstellar-tab}). 

The median planet radius does not vary significantly between the three
stellar groups ($2.2 - 2.6 R_\oplus$; see Table~\ref{fracstellar-tab}), but
the size of the largest planets in each group does.  The G and F stars
($T_{eff} > 5000$ K) contain Saturn-sized and Jupiter-sized planets (in
addition to smaller planets), while the cooler ($T_{eff} < 5000$ K) contain
Neptune-sized planets and smaller (see Figure~\ref{radiiteff-fig} and
Table~\ref{fracstellar-tab}).  In this sample, 15\% of the planets around
G-stars, and 25\% of the planets around the F-stars, are Neptune-sized or
larger; the cooler stars host only 6 (4\%) planets larger than Neptune, and
these four planets orbit three stars with effective temperatures of
$>4900$K  (KOI 757, 884, 941).  The absence of large planets  around cooler
stars is evident in the cumulative distribution of the planet radii
(Fig.~\ref{radiiteff-fig}) and in the smaller median absolute deviations of
the median planet radius for each stellar group
(Table~\ref{fracstellar-tab}).

This paper is not intended to provide a discussion of the occurrence rates
of planets or a detailed analysis of the size distribution of planets
\citep[e.g.,][]{howard12}, but this may imply that the cooler (i.e.,
smaller) stars only produce smaller planets that do not preferentially
follow a size hierarchy of planets .  In fact, large planets around small
stars may not form at all \citep{endl03, johnson10} or, if they do, they
may not survive their youth \citep[e.g.,][]{vaneyken12}. Thus, cool stars
may only be left with a population of relatively small planets
\citep{lfm12}.  The warmer stars, in contrast, appear to host a larger
array of planet sizes with a slightly higher preference for the larger
planets ($>$Neptune-sized) to be in longer orbits exterior to the orbits of
the smaller planets. Given that there appears to be only a weak 
correlation with the planet-radii ratio and the orbital period (ratio),
perhaps the formation and migration of large planets (perhaps in
conjunction with photo-evaporation of the innermost planets) is necessary
for a radius hierarchy to present.

\subsection{Planet Radius}\label{radius-sec}

\begin{figure}[htp]
\centering%
    \includegraphics[angle=0,scale=0.5,keepaspectratio=true]{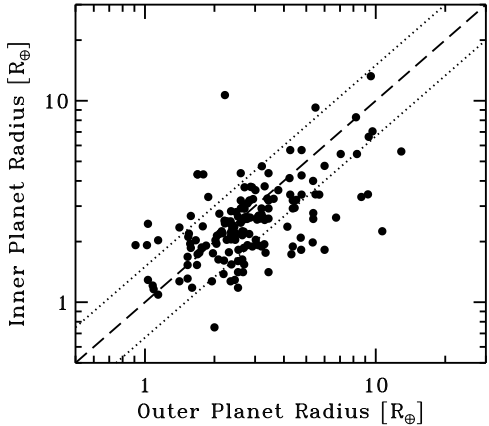}

    \figcaption{Comparison of the inner planet radius to the outer planet
    radius for each pair of planets, The dashed line delineates unity and
    the dotted lines mark the boundaries where one planet is 50\% bigger
    than the planet to which it is compared (ratio = 0.67 if the outer
    planet is larger and ratio = 1.5 if the inner planet is
    larger).\label{largeratio-fig}}
\end{figure}

If hotter stars tend to show a planet radius hierarchy and the hotter stars
also host relatively larger planets, then the planet hierarchy might be
expected to be correlated with the planet size.  In
Figure~\ref{largeratio-fig}, we compare the sizes of the planet radii for
the inner and outer planets for each pair of planets within the  sample. 
The overall planet-radii ratio is near unity but there is more scatter
below than above the unity line (i.e., smaller planets are interior to
larger planets).  There are 36 planet-pairs where the outer planet is
$>50\%$ larger than the inner planet ($R_{inner}/R_{outer} \le 0.67$); in
contrast, there are only 13 planet-pairs where the inner planet is $>50\%$
larger than the outer planet ($ R_{inner}/R_{outer} \ge 1.5)$.
\begin{figure}[h]
	\centering%
    \includegraphics[angle=0,scale=0.5,keepaspectratio=true]{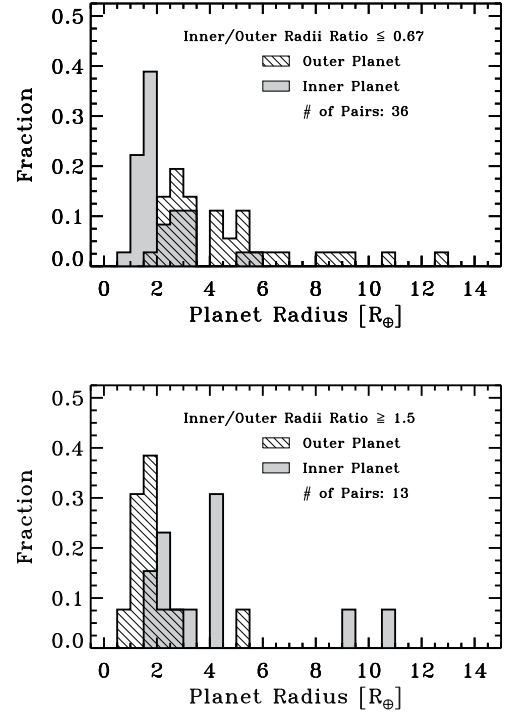}

    \figcaption{Distribution of the planet radii for those planet
    pairs where one planet is $>50\%$ larger than the other planet.
    The top panel is for those planet pairs where the outer planet is
    larger than the inner planet ($R_{inner}/R_{outer} \le 0.67$);
    the bottom panel is for those planet pairs where the inner planet is
    larger than the outer planet ($R_{inner}/R_{outer} \ge 1.5$);
    \label{largeratiohist-fig}}

\end{figure}

For the planet-pairs that are in these relatively extreme ratio pairs, we
have plotted the radii distributions of the candidates, separated out by
inner and  outer planet and by the sense of the ratio (see
Figure~\ref{largeratiohist-fig}). For the pairs where the outer planet is
significantly larger, the ratio tends to be dominated by relatively large
planets. Half (18/36) of the outer planets are Neptune-sized or larger
($>4R_{\oplus}$), while only 25\% (9/36) of the inner planets are very
small (e.g., Earth-sized $<1.5R_{\oplus}$) planets. In contrast, for those
planet-pairs where the inner planet is much larger than the outer planet,
the pairs are evenly split between relatively small outer planets (6/13 are
$<1.5R_{\oplus}$) and relatively large inner planets (5/13 $>4R_{\oplus}$).

\begin{figure}[htp]
\centering%
    \includegraphics[angle=0,scale=0.5,keepaspectratio=true]{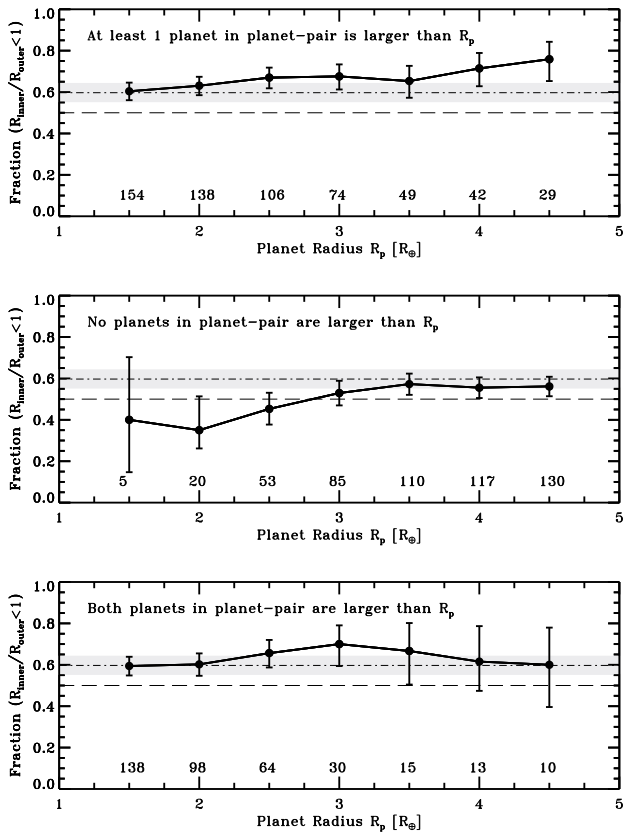}

    \figcaption{The fraction of planet-pairs with $(R_{inner}/R_{outer}<1)$
    are plotted as a function of maximum planet radius. In the top panel,
    only those planet-pairs are included that have at least one planet that
    is larger than a planet radius $R_p$.  In the middle panel, only those
    planet-pairs are included where both planets are smaller than a planet
    radius $R_p$.  In the bottom panel, only those planet-pairs are
    included where both planets are larger than a planet radius $R_p$.  The
    numbers underneath each data point list the number of pairs in that
    bin.  In each panel, the dashed-dot line and grey area delineate the
    fraction and uncertainties found for the entire sample (from
    Fig.~\ref{ratiocfrac-fig});  the dashed line delineates the 50\%
    fraction. \label{fracrad-fig}}

\end{figure}

It appears that for there to be a planet radius hierarchy, one of the
planets most often needs to be $\sim$Neptune-sized or larger.  To explore
this more closely, we  have calculated the fraction of planet-pairs with
$(R_{inner}/R_{outer}<1)$, if at least one planet is larger than some
maximum planet radius ($R_p$), or if both planets are smaller than that
same maximum planet radius ($R_p$), or if both planets are larger than that
same maximum planet radius.  The fractions were calculated for maximum
planet radii of $R_p = [1.5,2.0,2.5,3.0,3.5,4.0,4.5]R_\oplus$  (see
Figure~\ref{fracrad-fig}) and are tabulated in Table~\ref{fracrp-tab}. 

If planets-pairs with planets larger than $3R_\oplus$ are excluded, the
fraction is very near 50\% with little preference for sequential planet
ordering,  but if one planet in the pair is larger than $\sim3R_\oplus$,
the observed fraction of planet-pairs with $(R_{inner}/R_{outer}<1)$ is
$68\pm6\%$ -- a $3\sigma$ separation from the random ordering fraction of
50\% (see Table~\ref{fracrp-tab}). Based upon the $\chi^2$ statistic and
probability and the confidence intervals, the fraction of planet pairs is
significantly above 50\% only when Neptune-sized planets or larger are
allowed in the pairs.  

The fraction is most significant when Neptune-sized or larger planets are
compared to planets of all sizes (top panel Fig~\ref{fracrad-fig}); the
$\chi^2$ statistic are all $>4$ with a $\lesssim 1\%$ probability that the
non$-50\%$ fractions are achieved solely by chance.    When the sizes of
both planets are restricted to radii smaller than Neptune, the fractions
remain near or below 50\% (middle panel Fig~\ref{fracrad-fig} and
Table~\ref{fracrp-tab}), with a $\gtrsim 15\%$  probability of the observed
fractions being generated by chance for the null-hypothesis distribution. 
When both planets within a planet pair are restricted to planets larger
than a certain radius, sequential ordering of the planet sizes is still
apparent, but is less significant (bottom panel Fig~\ref{fracrad-fig} and
Table~\ref{fracrp-tab}).  If both planets in a pair are $\gtrsim
3.0R_\oplus$,  then the observed fraction of planet-pairs where the inner
planet is smaller than the outer planet are consistent with no planet size
hierarchy. These results suggest that the size ordering primarily occurs
when a system contains both superearth-sized (or earth-sized) planets {\it
and} Neptune-sized or larger planets, and may be a direct result
shepherding of smaller inner planets by larger outer planets
\citep{raymond08}.

\section{Summary}\label{summary-sec}

We have performed a study of the relative sizes of planets within the
multiple candidate systems using the $Kepler$ planetary candidate list
\citep{batalha12}.   For each multiple planet system, we have compared the
radius of each planet to the radius of every other planet within a
planetary system, after correcting for detection biases.  We find that for
planet-pairs for which one or both objects is  approximately Neptune-sized
for larger, the larger planet is most often the planet with the longer
period, while no such size--location  correlation is seen for pairs of
planets when both planets are smaller than Neptune.  Overall, for all
planets-pairs in the sample, $60\pm 4\%$ of the unique planet pairs are
structured in a hierarchical manner such that the inner planet is smaller
than the outer planet.  If at least one planet in the planet-pair is
Neptune-sized or larger ($\gtrsim 3 R_\oplus$), the fraction of inner
planets being smaller than outer planets is $\approx 68\pm 6\%$. However,
if both planets are smaller than Neptune, then the fraction is consistent
with random planet ordering ($53\pm6$\%) with no apparent size hierarchy. 

The planet radius size hierarchy may be a natural consequence of planetary
formation and evolution.  In particular, the sequential ordering may be a
result of a combination of core accretion, migration, and evolution.  Core
accretion models predict that smaller planets are expected to form prior to
and interior to the giant planets \citep{zhou05}.  Additionally, larger
planets formed further out may migrate inwards, shepherding smaller planets
inward as the planets move \citep{raymond08}.  

This scenario is consistent with the planet size hierarchy being observed
for planets pairs involving Neptune-sized planets, but being absent for
small planet pairs and for the planets around cool stars.  For the cooler
stars, the forming planets may begin migrating prior to starting rapid gas
accretion \citep[e.g.,][]{il05}, coupled with a lower extreme ultraviolet
luminosity that is not capable of substantially evaporating the planets
\citep{lfm12}. For example, \cite{lfm12} suggest that the $Kepler$-11
planets did not form {\it in situ},  but rather, the planets formed beyond
the snow line, migrated inwards, and were evaporated by the star to their
present sizes. 

These scenarios (perhaps, all in concert) predict a radius hierarchy of
planets as a function of orbital distance from the central host star, in
particular, predicting that Neptune-sized and larger planets are outside
super-Earth and terrestrial-sized planets. As $Kepler$ discovers more
planets in longer orbital periods, it will be interesting to learn if our
Solar System is indeed typical of the planetary architectures found in the
Galaxy.

\acknowledgments

$Kepler$ was competitively selected as the 10th NASA Discovery mission. The
authors thank the many people who have made $Kepler$ such a success.  This
paper includes data collected by the $Kepler$ mission; funding for the
$Kepler$ mission is provided by the NASA Science Mission directorate.This
research has made use of the NASA Exoplanet Archive, which is operated by
the California Institute of Technology, under contract with the National
Aeronautics and Space Administration under the Exoplanet Exploration
Program.  D.C.F. acknowledges NASA support through Hubble Fellowship grant
HF-51272.01-A, awarded by STScI, operated by AURA under contract NAS
5-26555.   DRC would like to thank the referee, the $Kepler$ team, Bill
Borucki, Geoff Marcy, Stephen Kane, Peter Plavchan, Kaspar von Braun,
Teresa Ciardi, and Jim Grubbs for very insightful and inspirational
comments and discussions in the formation of this paper.

\LongTables
\begin{deluxetable}{ccc|cccc|cccc|c}
\tablecolumns{12}
\tablewidth{8.25in}
\tablecaption{Summary of Planet-Radii Ratios\label{sum-tab}}
\tablehead{
\colhead{KOI} & 
\colhead{Stellar} & 
\colhead{Planet} &
\multicolumn{4}{|c|}{Inner Planet} &
\multicolumn{4}{c|}{Outer Planet} &
\colhead{Planet Radii}\\
\colhead{} & 
\colhead{Temp} &  
\colhead{Pair} &
\colhead{Radius} &
\colhead{Period} &
\colhead{Transit} &
\colhead{Impact} &
\colhead{Radius} &
\colhead{Period} &
\colhead{Transit} &
\colhead{Impact} &
\colhead{Ratio}\\
\colhead{} & 
\colhead{(K)} &  
\colhead{} &
\colhead{$(R_\oplus)$} &
\colhead{(days)} &
\colhead{SNR} &
\colhead{Par. }&
\colhead{$(R_\oplus)$} &
\colhead{(days)} &
\colhead{SNR} &
\colhead{Par.}&
\colhead{$R_{inner}/R_{outer}$}
}
\startdata
\multicolumn{12}{c}{KOIs with 2 Candidates}\\
K00072  &  5627  &    I/II  &   1.38  &    0.837  &   139  &  0.17  &   2.19  &   45.294  &   122  &  0.11  &  0.630\\
K00108  &  5975  &    I/II  &   2.94  &   15.965  &    84  &  0.79  &   4.45  &  179.600  &   111  &  0.62  &  0.661\\
K00119  &  5380  &    I/II  &   3.76  &   49.184  &   152  &  0.55  &   3.30  &  190.310  &    63  &  0.69  &  1.139\\
K00123  &  5871  &    I/II  &   2.64  &    6.482  &    98  &  0.56  &   2.71  &   21.222  &    78  &  0.01  &  0.974\\
K00150  &  5538  &    I/II  &   2.63  &    8.409  &   106  &  0.73  &   2.73  &   28.574  &    81  &  0.76  &  0.963\\
K00209  &  6221  &    I/II  &   5.44  &   18.795  &   217  &  0.51  &   8.29  &   50.790  &   287  &  0.40  &  0.656\\
K00222  &  4353  &    I/II  &   2.03  &    6.312  &    85  &  0.71  &   1.66  &   12.794  &    43  &  0.72  &  1.223\\
K00223  &  5128  &    I/II  &   2.52  &    3.177  &    90  &  0.78  &   2.27  &   41.008  &    32  &  0.69  &  1.110\\
K00271  &  6169  &    I/II  &   2.48  &   29.392  &    72  &  0.71  &   2.60  &   48.630  &    54  &  0.79  &  0.954\\
K00275  &  5795  &    I/II  &   1.95  &   15.791  &    55  &  0.03  &   2.04  &   82.199  &    27  &  0.60  &  0.956\\
K00312  &  6158  &    I/II  &   1.91  &   11.578  &    41  &  0.70  &   1.84  &   16.399  &    40  &  0.58  &  1.038\\
K00313  &  5348  &    I/II  &   1.61  &    8.436  &    54  &  0.32  &   2.20  &   18.735  &    55  &  0.73  &  0.732\\
K00386  &  5969  &    I/II  &   3.25  &   31.158  &    57  &  0.02  &   2.94  &   76.732  &    28  &  0.55  &  1.105\\
K00413  &  5236  &    I/II  &   2.75  &   15.228  &    52  &  0.67  &   2.10  &   24.674  &    26  &  0.65  &  1.310\\
K00431  &  5249  &    I/II  &   2.80  &   18.870  &    60  &  0.60  &   2.48  &   46.901  &    34  &  0.68  &  1.129\\
K00433  &  5237  &    I/II  &   5.60  &    4.030  &   127  &  0.65  &  12.90  &  328.240  &   158  &  0.77  &  0.434\\
K00446  &  4492  &    I/II  &   1.76  &   16.709  &    33  &  0.62  &   1.72  &   28.551  &    34  &  0.05  &  1.023\\
K00448  &  4264  &    I/II  &   1.77  &   10.139  &    56  &  0.74  &   2.31  &   43.608  &    66  &  0.66  &  0.766\\
K00464  &  5362  &    I/II  &   2.63  &    5.350  &    75  &  0.58  &   6.73  &   58.362  &   263  &  0.20  &  0.391\\
K00475  &  5056  &    I/II  &   2.04  &    8.181  &    37  &  0.58  &   2.26  &   15.313  &    34  &  0.68  &  0.903\\
K00509  &  5437  &    I/II  &   2.24  &    4.167  &    45  &  0.09  &   2.68  &   11.463  &    32  &  0.74  &  0.836\\
K00518  &  4565  &    I/II  &   2.11  &   13.981  &    68  &  0.71  &   1.54  &   44.000  &    33  &  0.49  &  1.370\\
K00638  &  5722  &    I/II  &   3.60  &   23.636  &    53  &  0.02  &   3.78  &   67.093  &    39  &  0.27  &  0.952\\
K00657  &  4632  &    I/II  &   1.63  &    4.069  &    43  &  0.63  &   2.08  &   16.282  &    46  &  0.74  &  0.784\\
K00672  &  5524  &    I/II  &   2.60  &   16.087  &    43  &  0.71  &   3.15  &   41.749  &    70  &  0.11  &  0.825\\
K00676  &  4367  &    I/II  &   2.56  &    2.453  &   255  &  0.70  &   3.30  &    7.972  &   348  &  0.57  &  0.776\\
K00693  &  6121  &    I/II  &   1.87  &   15.660  &    48  &  0.50  &   1.76  &   28.779  &    23  &  0.64  &  1.062\\
K00708  &  6036  &    I/II  &   1.82  &    7.693  &    43  &  0.74  &   2.54  &   17.406  &    68  &  0.70  &  0.717\\
K00800  &  5938  &    I/II  &   3.27  &    2.711  &    48  &  0.76  &   3.39  &    7.212  &    36  &  0.77  &  0.965\\
K00841  &  5399  &    I/II  &   5.44  &   15.334  &    83  &  0.74  &   7.05  &   31.331  &   117  &  0.74  &  0.772\\
K00842  &  4497  &    I/II  &   2.04  &   12.718  &    43  &  0.31  &   2.46  &   36.065  &    44  &  0.42  &  0.829\\
K00853  &  4842  &    I/II  &   2.38  &    8.204  &    48  &  0.12  &   1.78  &   14.496  &    22  &  0.02  &  1.337\\
K00870  &  4590  &    I/II  &   2.52  &    5.912  &    56  &  0.76  &   2.35  &    8.986  &    47  &  0.73  &  1.072\\
K00877  &  4500  &    I/II  &   2.42  &    5.955  &    54  &  0.73  &   2.37  &   12.039  &    36  &  0.79  &  1.021\\
K00896  &  5190  &    I/II  &   3.22  &    6.308  &    68  &  0.52  &   4.52  &   16.239  &    91  &  0.57  &  0.712\\
K00936  &  3684  &    I/II  &   1.54  &    0.893  &    72  &  0.75  &   2.69  &    9.468  &    96  &  0.76  &  0.572\\
K00951  &  4767  &    I/II  &   3.74  &   13.197  &   101  &  0.39  &   2.87  &   33.653  &    36  &  0.58  &  1.303\\
K00988  &  5218  &    I/II  &   2.20  &   10.381  &    67  &  0.02  &   2.17  &   24.570  &    37  &  0.67  &  1.014\\
K01236  &  6562  &    I/II  &   2.04  &   12.309  &    40  &  0.01  &   3.02  &   35.743  &    64  &  0.37  &  0.675\\
K01270  &  5145  &    I/II  &   2.19  &    5.729  &    45  &  0.58  &   1.55  &   11.609  &    22  &  0.01  &  1.413\\
K01781  &  4977  &    I/II  &   1.94  &    3.005  &    81  &  0.19  &   3.29  &    7.834  &   172  &  0.01  &  0.590\\
K01824  &  5978  &    I/II  &   1.75  &    1.678  &    53  &  0.70  &   1.97  &    3.554  &    50  &  0.69  &  0.888\\

\multicolumn{12}{c}{KOIs with 3 Candidates}\\
K00085  &  6172  &    I/II  &   1.27  &    2.155  &    72  &  0.71  &   2.35  &    5.860  &   154  &  0.77  &  0.540\\
K00085  &  6172  &   I/III  &   1.27  &    2.155  &    72  &  0.71  &   1.41  &    8.131  &    53  &  0.78  &  0.901\\
K00085  &  6172  &  II/III  &   2.35  &    5.860  &   154  &  0.77  &   1.41  &    8.131  &    53  &  0.78  &  1.667\\
K00111  &  5711  &    I/II  &   2.14  &   11.427  &    96  &  0.53  &   2.05  &   23.668  &    71  &  0.55  &  1.044\\
K00111  &  5711  &   I/III  &   2.14  &   11.427  &    96  &  0.53  &   2.36  &   51.756  &    75  &  0.57  &  0.907\\
K00111  &  5711  &  II/III  &   2.05  &   23.668  &    71  &  0.55  &   2.36  &   51.756  &    75  &  0.57  &  0.869\\
K00115  &  6202  &  II/III  &   3.33  &    5.412  &   144  &  0.43  &   1.88  &    7.126  &    41  &  0.55  &  1.771\\
K00137  &  5385  &    I/II  &   1.82  &    3.505  &    44  &  0.71  &   4.75  &    7.641  &   299  &  0.50  &  0.383\\
K00137  &  5385  &   I/III  &   1.82  &    3.505  &    44  &  0.71  &   6.00  &   14.858  &   305  &  0.72  &  0.303\\
K00137  &  5385  &  II/III  &   4.75  &    7.641  &   299  &  0.50  &   6.00  &   14.858  &   305  &  0.72  &  0.792\\
K00152  &  6187  &    I/II  &   2.59  &   13.484  &    58  &  0.34  &   2.77  &   27.402  &    57  &  0.07  &  0.935\\
K00152  &  6187  &   I/III  &   2.59  &   13.484  &    58  &  0.34  &   5.36  &   52.091  &   158  &  0.00  &  0.483\\
K00152  &  6187  &  II/III  &   2.77  &   27.402  &    57  &  0.07  &   5.36  &   52.091  &   158  &  0.00  &  0.517\\
K00156  &  4619  &    I/II  &   1.18  &    5.188  &    38  &  0.53  &   1.60  &    8.041  &    54  &  0.64  &  0.738\\
K00156  &  4619  &   I/III  &   1.18  &    5.188  &    38  &  0.53  &   2.53  &   11.776  &   131  &  0.69  &  0.466\\
K00156  &  4619  &  II/III  &   1.60  &    8.041  &    54  &  0.64  &   2.53  &   11.776  &   131  &  0.69  &  0.632\\
K00284  &  5925  &    I/II  &   1.09  &    6.178  &    32  &  0.38  &   1.14  &    6.415  &    35  &  0.19  &  0.956\\
K00343  &  5744  &    I/II  &   1.86  &    2.024  &    72  &  0.63  &   2.68  &    4.762  &   108  &  0.63  &  0.694\\
K00343  &  5744  &   I/III  &   1.86  &    2.024  &    72  &  0.63  &   1.58  &   41.809  &    20  &  0.53  &  1.177\\
K00343  &  5744  &  II/III  &   2.68  &    4.762  &   108  &  0.63  &   1.58  &   41.809  &    20  &  0.53  &  1.696\\
K00351  &  6103  &  II/III  &   6.62  &  210.590  &   144  &  0.22  &   9.32  &  331.640  &   211  &  0.19  &  0.710\\
K00377  &  5777  &  II/III  &   8.28  &   19.273  &   136  &  0.35  &   8.21  &   38.907  &    81  &  0.62  &  1.009\\
K00398  &  5101  &    I/II  &   1.76  &    1.729  &    37  &  0.24  &   3.33  &    4.180  &   102  &  0.00  &  0.529\\
K00398  &  5101  &  II/III  &   3.33  &    4.180  &   102  &  0.00  &   8.66  &   51.846  &   215  &  0.66  &  0.385\\
K00408  &  5631  &    I/II  &   3.72  &    7.382  &    96  &  0.79  &   2.91  &   12.560  &    52  &  0.77  &  1.278\\
K00408  &  5631  &   I/III  &   3.72  &    7.382  &    96  &  0.79  &   2.70  &   30.827  &    34  &  0.79  &  1.378\\
K00408  &  5631  &  II/III  &   2.91  &   12.560  &    52  &  0.77  &   2.70  &   30.827  &    34  &  0.79  &  1.078\\
K00481  &  5227  &    I/II  &   1.54  &    1.554  &    43  &  0.71  &   2.37  &    7.650  &    56  &  0.76  &  0.650\\
K00481  &  5227  &  II/III  &   2.37  &    7.650  &    56  &  0.76  &   2.44  &   34.260  &    39  &  0.73  &  0.971\\
K00528  &  5448  &   I/III  &   2.62  &    9.577  &    55  &  0.48  &   3.27  &   96.671  &    32  &  0.71  &  0.801\\
K00620  &  5803  &   I/III  &   7.05  &   45.155  &   132  &  0.03  &   9.68  &  130.180  &   279  &  0.06  &  0.728\\
K00658  &  5676  &    I/II  &   2.03  &    3.163  &    68  &  0.53  &   2.02  &    5.371  &    52  &  0.65  &  1.005\\
K00658  &  5676  &   I/III  &   2.03  &    3.163  &    68  &  0.53  &   1.14  &   11.329  &    17  &  0.01  &  1.781\\
K00665  &  5864  &    I/II  &   1.16  &    1.612  &    34  &  0.69  &   1.09  &    3.072  &    27  &  0.64  &  1.064\\
K00701  &  4807  &    I/II  &   1.27  &    5.715  &    55  &  0.46  &   1.95  &   18.164  &    71  &  0.66  &  0.651\\
K00701  &  4807  &  II/III  &   1.95  &   18.164  &    71  &  0.66  &   1.57  &  122.390  &    34  &  0.29  &  1.242\\
K00711  &  5502  &  II/III  &   3.18  &   44.699  &    51  &  0.61  &   2.83  &  124.520  &    35  &  0.49  &  1.124\\
K00718  &  5788  &    I/II  &   2.57  &    4.585  &    67  &  0.34  &   3.06  &   22.714  &    58  &  0.29  &  0.840\\
K00718  &  5788  &   I/III  &   2.57  &    4.585  &    67  &  0.34  &   2.67  &   47.904  &    27  &  0.68  &  0.963\\
K00718  &  5788  &  II/III  &   3.06  &   22.714  &    58  &  0.29  &   2.67  &   47.904  &    27  &  0.68  &  1.146\\
K00723  &  5244  &   I/III  &   3.26  &    3.937  &    72  &  0.79  &   3.61  &   28.082  &    60  &  0.58  &  0.903\\
K00757  &  4956  &    I/II  &   2.09  &    6.253  &    34  &  0.00  &   4.73  &   16.068  &   120  &  0.25  &  0.442\\
K00757  &  4956  &  II/III  &   4.73  &   16.068  &   120  &  0.25  &   3.21  &   41.192  &    40  &  0.35  &  1.474\\
K00806  &  5461  &  II/III  &  13.24  &   60.322  &   366  &  0.37  &   9.52  &  143.210  &    69  &  0.34  &  1.391\\
K00864  &  5337  &   I/III  &   2.51  &    4.312  &    63  &  0.14  &   2.23  &   20.050  &    30  &  0.00  &  1.126\\
K00884  &  4931  &  II/III  &   4.13  &    9.439  &   143  &  0.48  &   4.23  &   20.476  &    73  &  0.70  &  0.976\\
K00898  &  4648  &    I/II  &   2.18  &    5.170  &    35  &  0.64  &   2.83  &    9.771  &    47  &  0.68  &  0.770\\
K00898  &  4648  &  II/III  &   2.83  &    9.771  &    47  &  0.68  &   2.36  &   20.089  &    28  &  0.59  &  1.199\\
K00921  &  5046  &  II/III  &   2.66  &   10.281  &    43  &  0.69  &   3.09  &   18.119  &    46  &  0.70  &  0.861\\
K00941  &  4998  &    I/II  &   2.37  &    2.383  &    38  &  0.53  &   4.14  &    6.582  &    99  &  0.03  &  0.572\\
K00961  &  4188  &    I/II  &   2.63  &    0.453  &   100  &  0.69  &   2.86  &    1.214  &    73  &  0.77  &  0.920\\
K01576  &  5445  &    I/II  &   3.20  &   10.415  &    64  &  0.68  &   2.84  &   13.084  &    50  &  0.65  &  1.127\\
K01835  &  5004  &    I/II  &   2.69  &    2.248  &    43  &  0.44  &   3.11  &    4.580  &    37  &  0.67  &  0.865\\
K01860  &  5708  &  II/III  &   2.44  &    6.319  &    46  &  0.46  &   2.36  &   12.209  &    36  &  0.35  &  1.034\\
K01867  &  3892  &    I/II  &   1.21  &    2.549  &    37  &  0.40  &   1.08  &    5.212  &    25  &  0.31  &  1.120\\

\multicolumn{12}{c}{KOIs with 4 Candidates}\\
K00094  &  6217  &    I/II  &   1.41  &    3.743  &    35  &  0.16  &   3.43  &   10.423  &    78  &  0.01  &  0.411\\
K00094  &  6217  &  II/III  &   3.43  &   10.423  &    78  &  0.01  &   9.25  &   22.343  &   455  &  0.30  &  0.371\\
K00094  &  6217  &   II/IV  &   3.43  &   10.423  &    78  &  0.01  &   5.48  &   54.319  &   206  &  0.38  &  0.626\\
K00094  &  6217  &  III/IV  &   9.25  &   22.343  &   455  &  0.30  &   5.48  &   54.319  &   206  &  0.38  &  1.688\\
K00191  &  5495  &  II/III  &   2.25  &    2.418  &    53  &  0.55  &  10.67  &   15.358  &   642  &  0.59  &  0.211\\
K00191  &  5495  &  III/IV  &  10.67  &   15.358  &   642  &  0.59  &   2.22  &   38.651  &    21  &  0.52  &  4.806\\
K00245  &  5288  &  II/III  &   0.75  &   21.301  &    49  &  0.76  &   2.00  &   39.792  &   282  &  0.79  &  0.375\\
K00571  &  3881  &    I/II  &   1.31  &    3.887  &    36  &  0.76  &   1.53  &    7.267  &    36  &  0.76  &  0.856\\
K00571  &  3881  &  II/III  &   1.53  &    7.267  &    36  &  0.76  &   1.68  &   13.343  &    36  &  0.79  &  0.911\\
K00571  &  3881  &   II/IV  &   1.53  &    7.267  &    36  &  0.76  &   1.53  &   22.407  &    27  &  0.78  &  1.000\\
K00571  &  3881  &  III/IV  &   1.68  &   13.343  &    36  &  0.79  &   1.53  &   22.407  &    27  &  0.78  &  1.098\\
K00720  &  5123  &    I/II  &   1.41  &    2.796  &    58  &  0.08  &   2.66  &    5.690  &   124  &  0.40  &  0.530\\
K00720  &  5123  &   I/III  &   1.41  &    2.796  &    58  &  0.08  &   2.53  &   10.041  &    87  &  0.64  &  0.557\\
K00720  &  5123  &  II/III  &   2.66  &    5.690  &   124  &  0.40  &   2.53  &   10.041  &    87  &  0.64  &  1.051\\
K00733  &  5038  &  II/III  &   2.54  &    5.925  &    60  &  0.14  &   2.21  &   11.349  &    35  &  0.31  &  1.149\\
K00812  &  4097  &   I/III  &   2.19  &    3.340  &    54  &  0.13  &   2.11  &   20.060  &    28  &  0.35  &  1.038\\
K00834  &  5614  &  III/IV  &   1.98  &   13.233  &    33  &  0.38  &   5.33  &   23.653  &   154  &  0.38  &  0.371\\
K00869  &  5085  &   II/IV  &   2.73  &    7.490  &    43  &  0.47  &   3.20  &   36.280  &    34  &  0.42  &  0.853\\
K00880  &  5512  &  III/IV  &   4.00  &   26.442  &    48  &  0.64  &   5.35  &   51.530  &   109  &  0.02  &  0.748\\
K00952  &  3911  &  II/III  &   2.25  &    5.901  &    47  &  0.72  &   2.15  &    8.752  &    35  &  0.77  &  1.047\\
K00952  &  3911  &   II/IV  &   2.25  &    5.901  &    47  &  0.72  &   2.64  &   22.780  &    38  &  0.77  &  0.852\\
K00952  &  3911  &  III/IV  &   2.15  &    8.752  &    35  &  0.77  &   2.64  &   22.780  &    38  &  0.77  &  0.814\\
K01557  &  4783  &   II/IV  &   3.60  &    3.296  &   122  &  0.52  &   3.01  &    9.653  &    70  &  0.30  &  1.196\\
K01567  &  5027  &  II/III  &   2.46  &    7.240  &    34  &  0.36  &   2.22  &   17.326  &    23  &  0.00  &  1.108\\
K01930  &  5897  &  II/III  &   2.21  &   13.726  &    38  &  0.60  &   2.14  &   24.310  &    31  &  0.62  &  1.033\\
K01930  &  5897  &   II/IV  &   2.21  &   13.726  &    38  &  0.60  &   2.46  &   44.431  &    29  &  0.66  &  0.898\\
K01930  &  5897  &  III/IV  &   2.14  &   24.310  &    31  &  0.62  &   2.46  &   44.431  &    29  &  0.66  &  0.870\\

\multicolumn{12}{c}{KOIs with 5 Candidates}\\
K00070  &  5443  &    I/II  &   1.92  &    3.696  &   134  &  0.60  &   0.91  &    6.098  &    23  &  0.66  &  2.110\\
K00070  &  5443  &   I/III  &   1.92  &    3.696  &   134  &  0.60  &   3.09  &   10.854  &   260  &  0.55  &  0.621\\
K00070  &  5443  &    I/IV  &   1.92  &    3.696  &   134  &  0.60  &   1.02  &   19.577  &    18  &  0.74  &  1.882\\
K00070  &  5443  &     I/V  &   1.92  &    3.696  &   134  &  0.60  &   2.78  &   77.611  &   103  &  0.55  &  0.691\\
K00070  &  5443  &   III/V  &   3.09  &   10.854  &   260  &  0.55  &   2.78  &   77.611  &   103  &  0.55  &  1.112\\
K00082  &  4908  &  III/IV  &   1.29  &   10.311  &    79  &  0.71  &   2.45  &   16.145  &   172  &  0.73  &  0.527\\
K00082  &  4908  &   III/V  &   1.29  &   10.311  &    79  &  0.71  &   1.03  &   27.453  &    28  &  0.79  &  1.252\\
K00082  &  4908  &    IV/V  &   2.45  &   16.145  &   172  &  0.73  &   1.03  &   27.453  &    28  &  0.79  &  2.379\\
K00232  &  5868  &    I/II  &   1.73  &    5.766  &    41  &  0.40  &   4.31  &   12.465  &   207  &  0.43  &  0.401\\
K00232  &  5868  &   I/III  &   1.73  &    5.766  &    41  &  0.40  &   1.69  &   21.587  &    21  &  0.67  &  1.024\\
K00232  &  5868  &  II/III  &   4.31  &   12.465  &   207  &  0.43  &   1.69  &   21.587  &    21  &  0.67  &  2.550\\
K00232  &  5868  &   II/IV  &   4.31  &   12.465  &   207  &  0.43  &   1.79  &   37.996  &    21  &  0.56  &  2.408\\
K00232  &  5868  &    II/V  &   4.31  &   12.465  &   207  &  0.43  &   1.69  &   56.255  &    19  &  0.38  &  2.550\\
K00500  &  4613  &  III/IV  &   1.63  &    4.645  &    30  &  0.72  &   2.64  &    7.053  &    59  &  0.71  &  0.617\\
K00500  &  4613  &    IV/V  &   2.64  &    7.053  &    59  &  0.71  &   2.79  &    9.522  &    55  &  0.79  &  0.946\\
K00707  &  5904  &  II/III  &   3.42  &   13.175  &    37  &  0.77  &   5.69  &   21.775  &    80  &  0.77  &  0.601\\
K00707  &  5904  &   II/IV  &   3.42  &   13.175  &    37  &  0.77  &   4.26  &   31.784  &    42  &  0.77  &  0.803\\
K00707  &  5904  &    II/V  &   3.42  &   13.175  &    37  &  0.77  &   4.77  &   41.029  &    51  &  0.77  &  0.717\\
K00707  &  5904  &  III/IV  &   5.69  &   21.775  &    80  &  0.77  &   4.26  &   31.784  &    42  &  0.77  &  1.336\\
K00707  &  5904  &   III/V  &   5.69  &   21.775  &    80  &  0.77  &   4.77  &   41.029  &    51  &  0.77  &  1.193\\
K00707  &  5904  &    IV/V  &   4.26  &   31.784  &    42  &  0.77  &   4.77  &   41.029  &    51  &  0.77  &  0.893\\
K01589  &  5755  &  II/III  &   2.23  &    8.726  &    30  &  0.72  &   2.36  &   12.882  &    27  &  0.79  &  0.945\\

\multicolumn{12}{c}{KOIs with 6 Candidates}\\
K00157  &  5685  &    I/II  &   1.89  &   10.304  &    38  &  0.34  &   2.92  &   13.024  &    67  &  0.35  &  0.647\\
K00157  &  5685  &   I/III  &   1.89  &   10.304  &    38  &  0.34  &   3.20  &   22.686  &    73  &  0.33  &  0.591\\
K00157  &  5685  &    I/IV  &   1.89  &   10.304  &    38  &  0.34  &   4.37  &   31.995  &    87  &  0.79  &  0.432\\
K00157  &  5685  &  II/III  &   2.92  &   13.024  &    67  &  0.35  &   3.20  &   22.686  &    73  &  0.33  &  0.913\\
K00157  &  5685  &   II/IV  &   2.92  &   13.024  &    67  &  0.35  &   4.37  &   31.995  &    87  &  0.79  &  0.668\\
K00157  &  5685  &    II/V  &   2.92  &   13.024  &    67  &  0.35  &   2.60  &   46.687  &    40  &  0.49  &  1.123\\
K00157  &  5685  &   II/VI  &   2.92  &   13.024  &    67  &  0.35  &   3.43  &  118.360  &    54  &  0.36  &  0.851\\
K00157  &  5685  &  III/IV  &   3.20  &   22.686  &    73  &  0.33  &   4.37  &   31.995  &    87  &  0.79  &  0.732\\
K00157  &  5685  &   III/V  &   3.20  &   22.686  &    73  &  0.33  &   2.60  &   46.687  &    40  &  0.49  &  1.231\\
K00157  &  5685  &  III/VI  &   3.20  &   22.686  &    73  &  0.33  &   3.43  &  118.360  &    54  &  0.36  &  0.933\\
K00157  &  5685  &    IV/V  &   4.37  &   31.995  &    87  &  0.79  &   2.60  &   46.687  &    40  &  0.49  &  1.681\\
K00157  &  5685  &   IV/VI  &   4.37  &   31.995  &    87  &  0.79  &   3.43  &  118.360  &    54  &  0.36  &  1.274\\
K00157  &  5685  &    V/VI  &   2.60  &   46.687  &    40  &  0.49  &   3.43  &  118.360  &    54  &  0.36  &  0.758

\enddata 
\end{deluxetable}

\begin{deluxetable}{lcccccc}
\tablecolumns{7}
\tablewidth{7in}
\tablecaption{Planet-Radii Ratios Summary Grouped by Number of KOIs\label{frac-tab}}
\tablehead{
\colhead{ } &
\colhead{All} & 
\colhead{2-KOI} & 
\colhead{3-KOI} &
\colhead{4-KOI} &
\colhead{5-KOI} &
\colhead{6-KOI}\\
\colhead{ } &
\colhead{Systems} & \colhead{Systems} & \colhead{Systems}&
\colhead{Systems} & \colhead{Systems} & \colhead{Systems}
}
\startdata
\# of Stellar Systems & 96 & 42 & 33 & 14 & 6 & 1\\
\# of $R_{inner}/R_{outer}$ Pairs & 159   & 42    & 55    & 27    & 22    & 13 \\
Number $R_{inner}/R_{outer} < 1$ & 95 & 26 & 33 & 16 & 11 & 9\\
Number $R_{inner}/R_{outer} > 1$ & 64 & 16 & 22 & 11 & 11 & 4\\
$\chi^2$ Statistic\tablenotemark{a} & 6.0 & 2.4 & 2.2 & 0.93 & 0.0 & 1.9\\
$\chi^2$ Probability\tablenotemark{a} & 0.014 & 0.13 & 0.14 & 0.34 & 1.0 & 0.16\\
Fraction $R_{inner}/R_{outer} < 1$ & 0.597 & 0.619 & 0.600 & 0.592 & 0.500 & 0.692\\
Lower $1\sigma$ Confidence & 0.042 & 0.089 & 0.076 & 0.115 & 0.126 & 0.179\\
Upper $1\sigma$ Confidence & 0.041 & 0.082 & 0.072 & 0.107 & 0.126 & 0.140
\enddata
\tablenotetext{a}{The $\chi^2$ statistic and probability is based upon comparison of
the observed fractions to the null hypothesis fractions of 50\%.}

\end{deluxetable}

\begin{deluxetable}{lccc}
\tablecolumns{4}
\tablewidth{5.5in}
\tablecaption{Planet-Radii Ratios Grouped by Stellar Temperature\label{fracstellar-tab}}
\tablehead{
\colhead{ } &
\colhead{$>5800$ K} & 
\colhead{$5000-5800$ K} & 
\colhead{$<5000$K} \\
\colhead{ } &
\colhead{Stars} & \colhead{Stars} & \colhead{Stars}
}
\startdata
\# of Stellar Systems & 22 & 47 & 27\\
\# of $R_{inner}/R_{outer}$ Pairs & 40  & 79  & 40\\
Number $R_{inner}/R_{outer} < 1$ & 26 & 47 & 22\\
Number $R_{inner}/R_{outer} > 1$ & 14 & 32 & 18\\
$\chi^2$ Statistic & 3.6 & 2.9 & 0.4\\
$\chi^2$ Probability & 0.06 & 0.09 & 0.50\\
Fraction $R_{inner}/R_{outer} < 1$ & 0.650 & 0.594 & 0.550\\ 
Lower $1\sigma$ Confidence & 0.091 & 0.062 & 0.091 \\ 
Upper $1\sigma$ Confidence & 0.082 & 0.060 & 0.088 \\
Median Planet Radius $[R_\oplus]$ & 2.59 & 2.63 & 2.19\\
Med. Abs. Dev. $[R_\oplus]$ & 1.49 & 1.18 & 0.61\\
Min. Planet Radius $[R_\oplus]$ & 1.09 & 0.75 & 1.03\\
Max. Planet Radius $[R_\oplus]$ & 9.68 & 13.2 & 4.73
\enddata
\end{deluxetable}

\begin{deluxetable}{lccccccc}
\footnotesize
\tablecolumns{8}
\tablewidth{6.5in}
\tablecaption{Planet-Radii Ratios Summary Group by Maximum Planet Radius\label{fracrp-tab}}
\tablehead{
\colhead{$R_p = $ } &
\colhead{$1.5R_\oplus$} & 
\colhead{$2.0R_\oplus$} & 
\colhead{$2.5R_\oplus$} &
\colhead{$3.0R_\oplus$} &
\colhead{$3.5R_\oplus$} &
\colhead{$4.0R_\oplus$} &
\colhead{$4.5R_\oplus$}
}
\startdata
\multicolumn{8}{c}{At least one planet in planet-pair is larger than $R_p$}\\
\# of Stellar Systems & 93 & 86 & 63 & 44 & 29 & 23 & 18\\
\# of $R_{inner}/R_{outer}$ Pairs & 154   & 138    & 106    & 74    & 49    & 42 & 29\\
Number $R_{inner}/R_{outer} < 1$ & 93 & 87 & 71 & 50 & 32 & 30 & 22\\
Number $R_{inner}/R_{outer} > 1$ & 61 & 51 & 35 & 24 & 17 & 12 &  7\\
$\chi^2$ Statistic & 6.7 & 9.4 & 12.2 & 9.1 & 4.6 & 7.7 & 7.8\\
$\chi^2$ Probability & 0.009 & 0.002 & 0.0005 & 0.003 & 0.03 & 0.005 & 0.005\\
Fraction $R_{inner}/R_{outer} < 1$ & 0.603 & 0.630 & 0.670 & 0.676 & 0.653 & 0.714 & 0.759\\
Lower $1\sigma$ Confidence & 0.043 & 0.045 & 0.052 & 0.063 & 0.081 & 0.086 & 0.105\\
Upper $1\sigma$ Confidence & 0.041 & 0.043 & 0.048 & 0.058 & 0.073 & 0.075 & 0.084\\
\multicolumn{8}{c}{No planets in planet-pair are larger than $R_p$}\\
\# of Stellar Systems & 5 & 15 & 38 & 58 & 73 & 78 & 82\\
\# of $R_{inner}/R_{outer}$ Pairs & 5   & 20    & 53    & 85    & 110    &117 & 130\\
Number $R_{inner}/R_{outer} < 1$ & 2 & 7 & 24 & 45 & 63 & 65 & 73\\
Number $R_{inner}/R_{outer} > 1$ & 3 & 13 & 29 & 40 & 47 & 52 & 57\\
$\chi^2$ Statistic & 0.0 & 1.8 & 0.47 & 0.29 & 2.3 & 1.4 & 1.9\\
$\chi^2$ Probability & 1.0 & 0.18 & 0.49 & 0.59 & 0.13 & 0.23 & 0.16\\
Fraction $R_{inner}/R_{outer} < 1$ & 0.400 & 0.350 & 0.453 & 0.529 & 0.572 & 0.555 & 0.561\\
Lower $1\sigma$ Confidence & 0.253 & 0.088 & 0.076 & 0.060 & 0.052 & 0.050 & 0.048\\
Upper $1\sigma$ Confidence & 0.302 & 0.164 & 0.078 & 0.059 & 0.051 & 0.049 & 0.046\\
\multicolumn{8}{c}{Both planets in planet-pair are larger than $R_p$}\\
\# of Stellar Systems & 89 & 67 & 42 & 21 & 13 & 11 & 10\\
\# of $R_{inner}/R_{outer}$ Pairs & 138   & 98    & 64    & 30    & 15    & 13 & 10\\
Number $R_{inner}/R_{outer} < 1$ & 82 & 59 & 42 & 21 & 10 & 8 & 6\\
Number $R_{inner}/R_{outer} > 1$ & 56 & 39 & 22 & 9 & 5 & 5 & 4\\
$\chi^2$ Statistic & 4.9 & 4.1 & 6.2 & 4.8 & 1.7 & 0.69 & 0.4\\
$\chi^2$ Probability & 0.03 & 0.04 & 0.01 & 0.03 & 0.20 & 0.40 & 0.50\\
Fraction $R_{inner}/R_{outer} < 1$ & 0.594 & 0.602 & 0.656 & 0.700 & 0.667 & 0.615 & 0.600\\
Lower $1\sigma$ Confidence & 0.046 & 0.055 & 0.069 & 0.106 & 0.162 & 0.141 & 0.204\\
Upper $1\sigma$ Confidence & 0.044 & 0.053 & 0.064 & 0.091 & 0.135 & 0.172 & 0.180
\enddata
\end{deluxetable}

\begin{thebibliography}{}

\bibitem[Batalha et al.(2011)]{batalha11} Batalha, N.~M. et al.\ 2011,
\apj, 729, 27

\bibitem[Batalha et al.(2012)]{batalha12} Batalha, N.~M., Rowe, 
J.~F., Bryson, S.~T., et al.\ 2012, arXiv:1202.5852 

\bibitem[Borucki et al.(2010)]{borucki10} Borucki, W.~J., et al.\
2010, Science, 327, 977

\bibitem[Borucki et al.(2011)]{borucki11} Borucki, W.~J., Koch, 
D.~G., Basri, G., et al.\ 2011, \apj, 736, 19

\bibitem[Ciardi et al.(2011)]{crd11} Ciardi, D. R. et al., 2011,
\aj, 141, 108  

\bibitem[Clopper \& Pearson(1934)]{cp34}Clopper, C.J., \&  Pearson, E.S.,
1934, Biometrika, 26, 404

\bibitem[Endl et al.(2003)]{endl03} Endl, M., Cochran, W.~D., 
Tull, R.~G., \& MacQueen, P.~J.\ 2003, \aj, 126, 3099  

\bibitem[Fabrycky et al.(2012a)]{fabrycky12a} Fabrycky, D.~C., Ford, 
E.~B., Steffen, J.~H., et al.\ 2012, \apj, 750, 114

\bibitem[Fabrycky et al.(2012b)]{fabrycky12b} Fabrycky, D.~C.et al.\ 2012, 
\apj, submitted, arXiv:1202.6328v2

\bibitem[Fang \& Margot(2012)]{fm12} Fang, J., \& Margot,
J.-L.\ 2012, arXiv:1207.5250 

\bibitem[Fressin et al.(2012)]{fressin12} Fressin, F., Torres, 
G., Rowe, J.~F., et al.\ 2012, \nat, 482, 195

\bibitem[Gautier et al.(2012)]{gautier12} Gautier, T.~N., III, 
Charbonneau, D., Rowe, J.~F., et al.\ 2012, \apj, 749, 15

\bibitem[Howard et al.(2012)]{howard12} Howard, A.~W., Marcy, 
G.~W., Bryson, S.~T., et al.\ 2012, \apjs, 201, 15 

\bibitem[Ida \& Lin(2005)]{il05} Ida, S. \& Lin, D. N. C. 2005, 
\apj, 626, 1045

\bibitem[Johnson et al.(2010)]{johnson10} Johnson, J.~A., Aller, 
K.~M., Howard, A.~W., \& Crepp, J.~R.\ 2010, \pasp, 122, 905 

\bibitem[Latham et al.(2011)]{latham11} Latham, D. W., et al.\ 2011.
\apjl, 732, L24

\bibitem[Lissauer et al.(2011a)]{lissauer11a} Lissauer, J.~J.,  Fabrycky,
D.~C., Ford, E.~B., et al.\ 2011, \nat, 470, 53 

\bibitem[Lissauer et al.(2011b)]{lissauer11b} Lissauer, J.~J., 
Ragozzine, D., Fabrycky, D.~C., et al.\ 2011, \apjs, 197, 8 

\bibitem[Lissauer et al.(2012)]{lissauer12} Lissauer, J.~J., 
Marcy, G.~W., Rowe, J.~F., et al.\ 2012, \apj, 750, 112

\bibitem[Lopez et al.(2012)]{lfm12} Lopez, E.~D., Fortney, 
J.~J., \& Miller, N.~K.\ 2012, arXiv:1205.0010

\bibitem[Migaszewski et al.(2012)]{msg12} Migaszewski, C., 
Slonina, M., \& Gozdziewski, K.\ 2012, arXiv:1205.0822
 
\bibitem[Morton \& Johnson(2011)]{mj11} Morton, T.~D., \& Johnson, J.~A.\
2011, \apj, 738, 170 

\bibitem[Muirhead et al.(2012)]{muirhead12} Muirhead, P.~S., 
Hamren, K., Schlawin, E., et al.\ 2012, \apjl, 750, L37 

\bibitem[Orosz et al.(2012)]{orosz12} Orosz, J.~A., Welsh, 
W.~F., Carter, J.~A., et al.\ 2012, Science, 337, 1511 

\bibitem[Press et al.(2007)]{nr07}Press, William H.; Teukolsky, Saul A.;
Vetterling, William T.; Flannery, Brian P. 2007, Numerical Recipes: The
Art of Scientific Computing (3rd ed.). New York: Cambridge University Press

\bibitem[Raymond et al.(2008)]{raymond08} Raymond, S.~N., Barnes, 
R., \& Mandell, A.~M.\ 2008, \mnras, 384, 663 

\bibitem[Raymond et  al.(2012)]{raymond12} Raymond, S.~N., Armitage, P.~J.,
Moro-Mart{\'{\i}}n, A., et al.\ 2012, \aap, 541, A11

\bibitem[Santerne et al.(2012)]{santerne12} Santerne, A., 
D{\'{\i}}az, R.~F., Moutou, C., et al.\ 2012, arXiv:1206.0601

\bibitem[Steffen et al.(2012a)]{steffen12a} Steffen, J.~H., 
Fabrycky, D.~C., Ford, E.~B., et al.\ 2012, \mnras, 421, 2342

\bibitem[Steffen et al.(2012b)]{steffen12b} Steffen, J.~H.,  Ragozzine, D.,
Fabrycky, D.~C., et al.\ 2012, Proceedings of the National  Academy of
Science, 109, 7982 (arXiv:1205.2309)

\bibitem[Torres et al.(2011)]{torres11} Torres, G. et al. \ 2011
\apj, 727, 24

\bibitem[Wright et al.(2009)]{wright09} Wright, J.~T., Upadhyay,  S.,
Marcy, G.~W., et al.\ 2009, \apj, 693, 1084 

\bibitem[van Eyken, Ciardi, von Braun et al.(2012)]{vaneyken12} van Eyken,
J.~C., Ciardi, D.~R., von Braun, K., et al.\ 2012, \apj, 755, 42 

\bibitem[von Braun, Kane, \& Ciardi(2009)]{vbkc09} von Braun, K., Kane, 
S.~R., \& Ciardi, D.~R.\ 2009, \apj, 702, 779 

\bibitem[Walsh et al.(2011)]{walsh11} Walsh, K.~J., Morbidelli, 
A., Raymond, S.~N., O'Brien, D.~P., \& Mandell, A.~M.\ 2011, \nat, 475, 206 

\bibitem[Zhou et al.(2005)]{zhou05} Zhou, J.-L., Aarseth,  S.~J., Lin,
D.~N.~C., \& Nagasawa, M.\ 2005, \apjl, 631, L85 

\end{thebibliography}
\end{document}